\documentclass[twoside,leqno,twocolumn]{article}
\usepackage{ltexpprt}
\usepackage{amsmath}
\usepackage{graphicx}
\usepackage{color}
\usepackage{algorithmic} % <- Preamble
\usepackage{algorithm}
\usepackage{caption}
\usepackage{subcaption}
\usepackage{cleveref}
\usepackage{multirow}
\usepackage{booktabs}
\Crefname{ALC@unique}{Line}{Lines} 

\begin{document}

%\setcounter{chapter}{2} % If you are doing your chapter as chapter one,
%\setcounter{section}{3} % comment these two lines out.

%\title{\Large Preserving Local and Global Information for Network Embedding\thanks{Supported by GSF grants ABC123, DEF456, and GHI789.}}
%\author{Yao Ma\thanks{Society for Industrial and Applied Mathematics.} \\
%\and
%Suhang Wang\thanks{Society for Industrial and Applied Mathematics.}\\
%\and
%Jiliang Tang}

\title{\Large Preserving Local and Global Information for Network Embedding}
\author{Yao Ma \thanks{Michigan State University, \{mayao4,tangjili\}@msu.edu}\\
\and
Suhang Wang \thanks{Arizona State University, suhang.wang@asu.edu}\\
\and
\and Zhaochun Ren\thanks{JD.com, \{renzhaochun,yindawei\}@jd.com}\\
\and Dawei Yin\footnotemark[3]\\
\and 
Jiliang Tang\footnotemark[1]}

\date{}

\maketitle

% Copyright Statement
% When submitting your final paper to a SIAM proceedings, it is requested that you include 
% the appropriate copyright in the footer of the paper.  The copyright added should be 
% consistent with the copyright selected on the copyright form submitted with the paper.
% Please note that "20XX" should be changed to the year of the meeting.

% Default Copyright Statement
\fancyfoot[R]{\footnotesize{\textbf{Copyright \textcopyright\ 20XX by SIAM\\
Unauthorized reproduction of this article is prohibited}}}

% Depending on which copyright you agree to when you sign the copyright form, the copyright 
% can be changed to one of the following after commenting out the default copyright statement
% above.

%\fancyfoot[R]{\footnotesize{\textbf{Copyright \textcopyright\ 20XX\\
%Copyright for this paper is retained by authors}}}

%\fancyfoot[R]{\footnotesize{\textbf{Copyright \textcopyright\ 20XX\\
%Copyright retained by principal author's organization}}}

%\pagenumbering{arabic}
%\setcounter{page}{1}%Leave this line commented out.

\begin{abstract} \small\baselineskip=9pt 
Networks such as social networks, airplane networks, and citation networks are ubiquitous. The adjacency matrix is often adopted to represent a network,  which is usually high dimensional and sparse. However, to apply advanced machine learning algorithms to network data, low-dimensional and continuous representations are desired. To achieve this goal, many network embedding methods have been proposed recently.  The majority of existing methods facilitate the local information i.e. local connections between nodes, to learn the representations, while completely neglecting global information (or node status), which has been proven to boost numerous network mining tasks such as link prediction and social recommendation. Hence, it also has potential to advance network embedding. In this paper, we study the problem of preserving local and global information for network embedding.  In particular, we introduce an approach to capture global information and propose a network embedding framework LOG, which can coherently model {\bf LO}cal and {\bf G}lobal information. Experimental results demonstrate the ability to preserve global information of the proposed framework. Further experiments are conducted to demonstrate the effectiveness of learned representations of the proposed framework.   
\end{abstract}
\section{Introduction}
% Information networks are ubiquitous in our life. Social networks, airplane networks, publication networks and the World Wide Web are some of the popular examples. The interaction between nodes of the information networks can be represented by adjacency matrices. However, it is very common that the number of nodes in the networks is large and the interaction between nodes is sparse, which, in turn, makes the adjacency matrix high dimensional and sparse. It is descried to find low-dimensional and continuous representations for information networks.
Networks, such as social networks, airplane networks, and citation networks, are ubiquitous and important in our daily life. Node representation learning/extraction is to learn or extract meaningful features for each node such that the representation can be facilitated in network analysis tasks such as link prediction, visualization, community detection and node classification. One naive way is to represent the nodes as an adjacency matrix, i.e., each node is represented as a sparse vector that contains link information. However, this representation usually suffers from the curse of dimensionality as the adjacency matrix is high dimensional and sparse because the networks are usually large with sparse links. Therefore, network embedding, which aims to learn low-dimensional and continuous node representations by preserving certain properties of the network, has attracted increasing attention in recent years~\cite{perozzi2014deepwalk,tang2015line,wang2017signed,grover2016node2vec}.     

% \begin{figure}[h!]
% %\centering
%  \centerline{\includegraphics[width=\linewidth]{figures/global_local}}
%  \caption{An illustration example of an network}
%  \label{fig:network}
% \end{figure}

\begin{figure}[!ht]
\centering 
  \begin{subfigure}[b]{0.45\linewidth}
  \centering
    \includegraphics[width=0.9\linewidth]{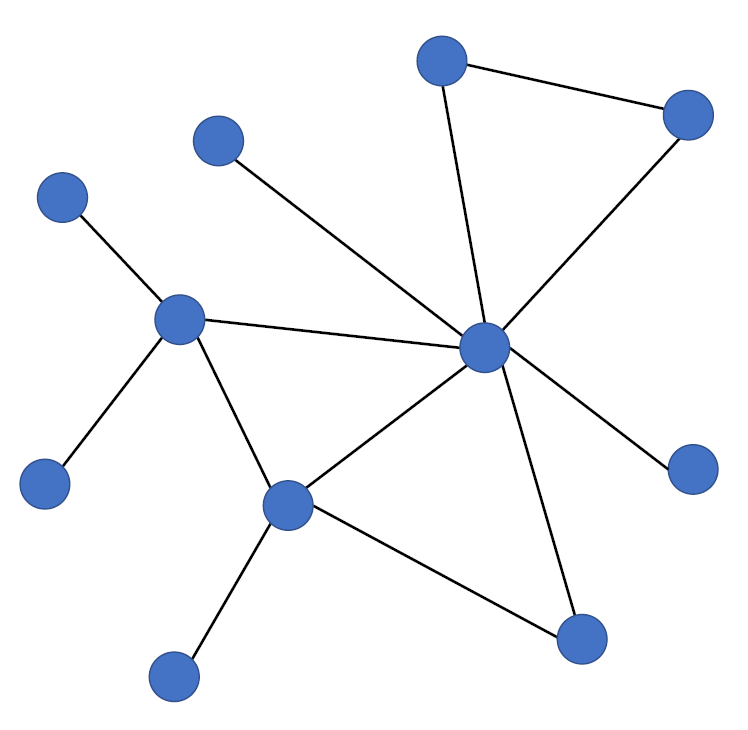}
    \caption{\footnotesize A Local view}
    \label{fig:local_view}
  \end{subfigure}
 \begin{subfigure}[b]{0.45\linewidth}
 \centering
    \includegraphics[width=0.9\linewidth]{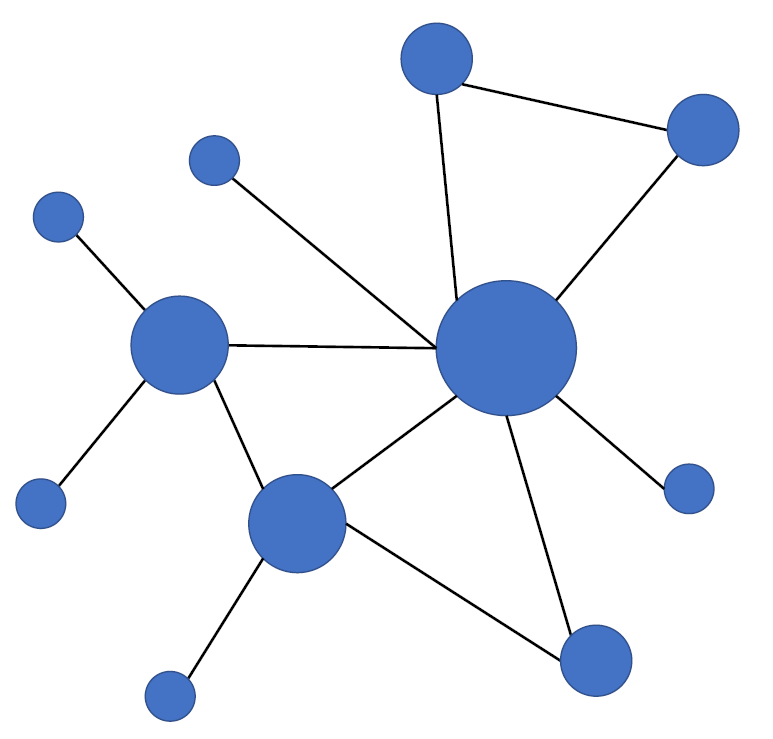}
    \caption{\footnotesize A Local and Global view}
    \label{fig:gobal_view}
  \end{subfigure}
  \caption{Viewing a network from different perspectives. }
  \label{fig:network}
 % \vspace{1mm}
\end{figure}
The majority of existing network embedding algorithms exploits local information to learn node representations. For example, DeepWalk~\cite{perozzi2014deepwalk} uses local information obtained from truncated random walks to learn latent representations; while LINE~\cite{tang2015line} preserves the first order proximity and second order proximity between nodes. By exploiting the local information, the learned vector representation of nodes captures certain local properties of the network, which has been demonstrated to advance many network analysis tasks such as link prediction~\cite{perozzi2014deepwalk,wang2017signed}, node classification~\cite{tang2015pte,wang2016linked} and visualization~\cite{tang2016visualizing}. 

Despite the local information, global information (or node status) is another easily accessible but important information. Node status, which reflects where a node stands in the entire network, has different meanings under different ``context''. For example, in the web networks, status can indicate a relevancy ranking of webpages~\cite{page1999pagerank}; in online social networks, it denotes reputations of users~\cite{massa2007survey}; and in healthcare and biological networks, it can help determine the node representativeness~\cite{ramakrishnan2009mining,tang2010identifing}. Most of the existing network embedding algorithms only facilitate local information by treating global status of each node equally as shown in Figure \ref{fig:local_view}. However, it is natural that nodes are of different global status as shown in Figure~\ref{fig:gobal_view}. Many network based tasks such as web search~\cite{page1999pagerank}, social network recommendation~\cite{tang2013exploiting}, active learning~\cite{hu2013actnet} and novel disease gene detection~\cite{lee2011prioritizing} have been proven to be advanced by exploiting the global status. Therefore, global status can provide complementary information in addition to the local structure and incorporating global information has great potential to help learn better representation. However, the work on exploiting both local and global information is rather limited.  

Therefore, in this paper, we study a new problem of investigating both local and global information for network representation learning. In essence, we need to solve two challenges: (1) How to mathematically capture global information; and (2) How to simultaneously model local and global information for network representation learning. In an attempt to solve these two challenges, we propose a novel framework LOG, which learns network representation that preserves both local and global information. The main contributions of the paper are summarized as follows:
\begin{itemize}
	\item We propose a principled way to model global information for network representation learning;
	\item We propose a novel network embedding framework LOG, which integrates local and global information into a coherent model;
	\item We conduct experiments on real-world datasets to demonstrate the effectiveness of the proposed framework.
\end{itemize}
The remaining of the paper is organized as follows. In Section \ref{sec:related_work}, we briefly review the related work. In Section \ref{sec:problem_statement}, we formally define the problem and then, we introduce the proposed framework in Section \ref{sec:model}. The experimental results and analysis are shown in Section \ref{sec:experiments}. We conclude our paper with discussion and future work in Section \ref{sec:conclusions}.
% Centrality measures which include pagerank [], degree centrality [], closeness centrality[] and betweeness centrality can be used to represent the global status of the nodes. So, we preserve these global ranking while learning the representations.
\section{Related Work}
\label{sec:related_work}
In this section, we briefly review some works related to our problem. Network embedding algorithms, which aim to learn low-dimensional node representation are attracting increasing attention recently. Inspired by word2vec~\cite{mikolov2013distributed,mikolov2013efficient}, DeepWalk~\cite{perozzi2014deepwalk} and LINE~\cite{tang2015line} are proposed. The general idea of word2vec is that a word can be explained by its surrounding neighbors. DeepWalk regards the nodes in a network as ``words'' of an ``artificial'' language and then use random walk to generate ``sentences'' for this language. Node representation can be learned following the procedure of word2vec. LINE tries to model the probability of the ``context'' of a node using the node representations and obtains the representations by minimizing the difference between the modeled probability and the empirical probability. node2vec\cite{grover2016node2vec} further extends DeepWalk by introducing parameters to allow biased random walk to explore the neighborhood of nodes. metapath2vec\cite{dong2017metapath2vec} extends the network embedding methods to heterogeneous network by introducing metapath based random walk. struc2vec\cite{ribeiro2017struc2vec} learns the node representation from a different perspective and it tries to preserve the structural identity between nodes. To achieve this goal, struc2vec first creates a new network based on the structural similarity between nodes and then follow the similar way of DeepWalk to learn node representation based on the created network. SDNE~\cite{wang2016structural} uses an auto-encoder to compress indicate vector (the row corresponding to the node in the adjacency matrix) to obtain low-dimensional representations. LANE~\cite{huang2017label} tries to learn node representations on the attributed network with label information. In \cite{wang2017signed}, a signed network embedding algorithm SiNE is proposed based on the notion that a user should be closer to their ``friend'' than their ``enemy''. SiNE uses a deep neural network to optimize this objective when sampling triplets from the signed network. Two recent surveys~\cite{hamilton2017representation,goyal2017graph} give a comprehensive overview of network embedding algorithms. However, most of these existing methods cannot preserve the global information. In this paper, we propose a model which can preserve the local information as well as the global information.

\section{Problem Statement}
\label{sec:problem_statement}
Before formally defining the problem we want to study, we first introduce the notations. Throughout the paper, matrices are written in boldface capital letters and vectors are denoted as boldface lowercase letters. 
%For an arbitrary matrix $\mathbf{M} \in \mathbf{R}^{m \times n}$, $\mathbf{M}_{ij}$ denotes the $(i,j)$-th entry of $\mathbf{M}$ while $\mathbf{m}^i$ and $\mathbf{m}_j$ mean the $i$-th row and $j$-th column of $\mathbf{M}$, respectively. $||\mathbf{M}||_F$ is the Frobenius norm of $\mathbf{M}$. 
Capital letters in calligraphic math font such as $\mathcal{P}$ are used to denote sets.  

Let $\mathcal{G = \{\mathcal{V},\mathcal{E}\}}$ be a network, where $\mathcal{V}=\{v_1,\dots,v_N\}$ denotes the set of $N$ nodes and $\mathcal{E}=\{e_1,\dots,e_M\}$ represents the set of $M$ edges between these nodes. Furthermore, let $\mathbf{t} \in  \mathbf{R}^{N \times 1}$ be the status scores to denote the global information for the $N$ nodes. The score can be calculated from various node status measures such as PageRank~\cite{page1999pagerank}. A larger score means a higher status.

With the aforementioned notations and definitions, the problem under study is formally stated as:

\emph{Given a network $\mathcal{G = \{\mathcal{V},\mathcal{E}\}}$ and  the global information $\mathbf{t}$, we aim to learn the node representation matrix $\mathbf{U} \in \mathbf{R}^{N \times d}$ by preserving the local structure information as well as the global information, where $d$ is the embedding dimension. Mathematically, the problem is written as:}
{\small
\vspace{-2mm}
\begin{align}
	Q(\mathcal{G}, \mathbf{t}) \rightarrow \mathbf{U}
\end{align}
}
\vspace{-2mm}
\emph{where $Q$ is the learning algorithm we will investigate.}

\section{The Proposed Framework}
\label{sec:model}
In this section, we introduce the proposed model. Our model consists of two components -- 1) the component to preserve the global information and 2) the component to preserve the local information. We first describe the component to preserve the global information, which leads to a new algorithm GINE. GINE can learn the embedding from only global information. We then introduce the component to preserve the local information and finally integrates both parts as the proposed framework LOG with both local and global information.

\subsection{Global Information Preserved Network Embedding.}
\vspace{-2mm}
In this subsection, we introduce a new embedding algorithm GINE which can preserve the global information for network embedding. Specifically, assume that we are given the status score vector $\mathbf{t}$ where $t_i$ is the status score for node $i$, which can be computed in various ways~\cite{page1999pagerank,okamoto2008ranking,newman2005measure}. In this work, we calculate the status scores based on Pagerank~\cite{page1999pagerank}. The status rankings have been widely used in real-world applications instead of the status scores; hence, we first get the status rankings of nodes based on their status scores and then directly preserve the status rankings for network embedding. 

To preserve the global ranking, we model this problem as a maximum likelihood problem.  In other words, we need to maximize the probability that the ranking of the learned statuses for the nodes $\{v_1,\dots,v_N\}$ follow the original Pagerank ranking $\{r_1,\dots,r_N\}$. For convenience, we use the ranking as the index for the nodes, that is, we use $\{v_{(r_1)},\dots,v_{(r_N)}\}$ to represent the nodes. Intuitively, if we can preserve the relative ranking of all pairs of nodes, we can maintain the whole ranking for all nodes. Thus, we assume that the order between a pair of nodes is independent of the other pairs, we approximate this probability as: 
{\small
\begin{align}\label{eq:p_global}
p_{global} = \prod\limits_{1\leq i <j\leq N} p(v_{(i)},v_{(j)});
\end{align}
}
where $p(v_{(i)},v_{(j)})$ is the probability that node $v_{(i)}$ is ranked before node $v_{(j)}$, which is defined as
{\small
\begin{align}
p(v_{(i)},v_{(j)}) = \sigma(f({\bf u}_i) - f({\bf u}_j)),
\end{align}
}
where $f(\cdot)$ is the mapping function that maps the node representation to the status and $\sigma(\cdot)$ is the sigmoid function
{\small
\begin{align}
\sigma(x) = \frac{1}{1+e^{-x}}.
\end{align}
}
Different mapping functions such as linear functions, non-linear functions and even neural networks can be chosen. In this paper, we adopt a linear function as
{\small
\begin{align}
f({\bf u}) = {\bf w}^T {\bf u}.
\end{align}
}
where $\mathbf{w}$ is the parameter of the function. We leave non-linear functions or neural networks as one future direction.

The representations and the mapping function $f(\cdot)$ can be learned by minimizing the negative logarithm of~\eqref{eq:p_global} as
{\small
\begin{align}
\label{eq:optimization_global}
 \mathbf{U},\mathbf{w} = &\arg\min\limits_{\mathbf{U},\mathbf{w}}  -\sum\limits_{1\leq i <j\leq N}\log(p(v_{(i)},v_{(j)})) \\
= & \arg\min\limits_{\mathbf{U},\mathbf{w}}  -\sum\limits_{1\leq i <j\leq N}\log (\sigma(f({\bf u}_i) - f({\bf u}_j))).\nonumber
\end{align}
}  
%The representations and the mapping function can be learned by minimizing the following objective function
%\begin{align}
%O_2 =& -\log P_{global} 
%\label{eq:objective2}
%\end{align}
\subsubsection{Reduce Computational Cost.}\label{sec:optimization_global}
\vspace{-2mm}
It is computational expensive to optimize~\eqref{eq:optimization_global} w.r.t $\mathbf{U}$ and $\mathbf{w}$ as there are $N(N-1)/2$ pairs in total. To accelerate the training speed, we relax the global ranking constraint. 

  \begin{figure}[h]
    \includegraphics[width=\linewidth]{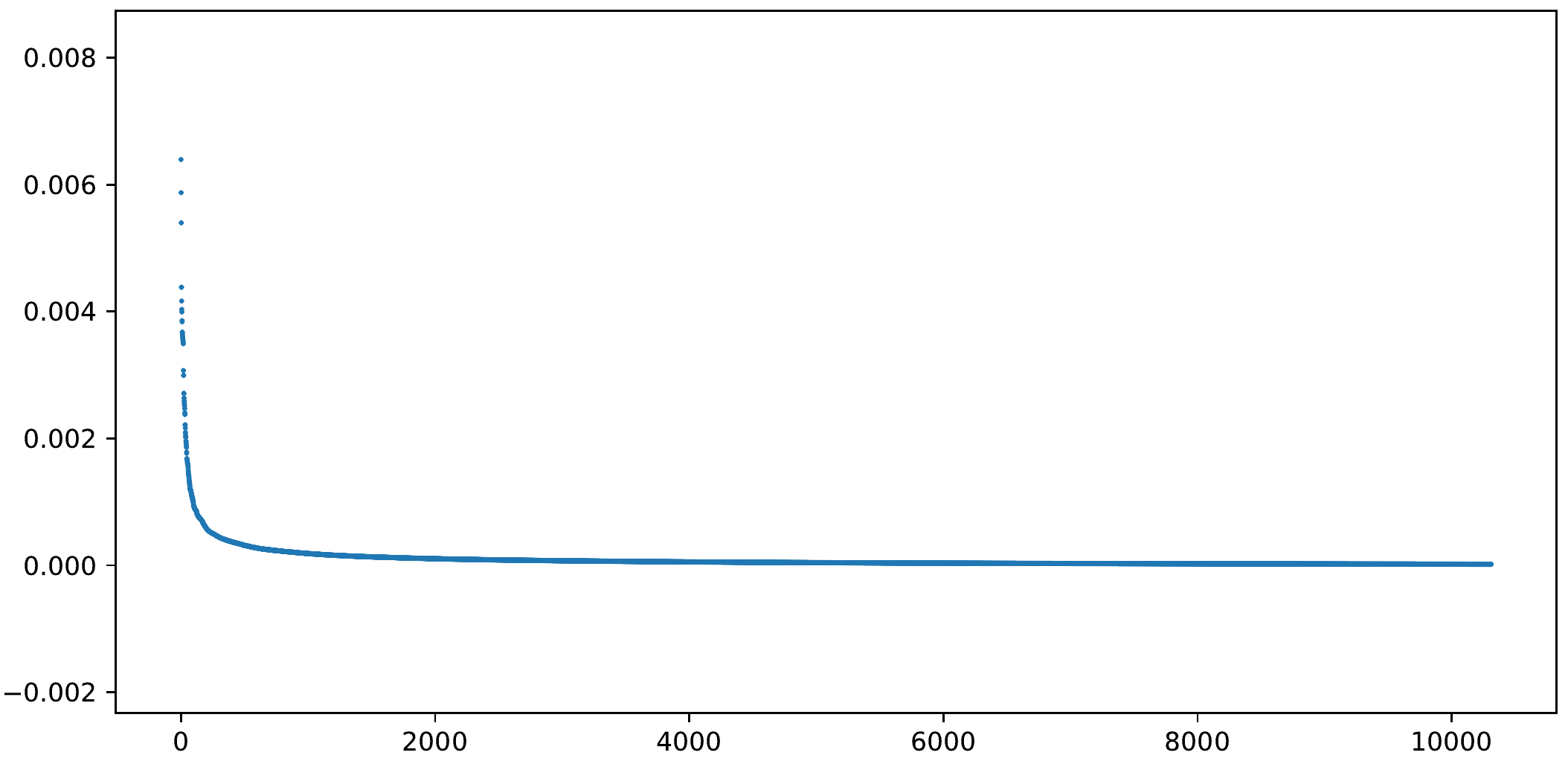}
    \caption{The descending ordered pagerank score of nodes in BlogCatalog network.}
    \label{fig:blog_pagerank}
  \end{figure}
\vspace{-2mm}
Our relaxation comes from the observation from the distribution of status scores. Figure~\ref{fig:blog_pagerank} plots the Pagerank value of the BlogCatalog network~\cite{Zafarani+Liu:2009} where $x$-axis is the number of nodes and $y$-axis is the PageRank score. We make two observations -- (1) many status scores are small; and (2) status scores in the long tail are very close to each other. Hence, the difference of the status scores of two nodes with different rankings in the long tail is very small. It is not useful to address the difference between these rankings. Therefore, we first transform the global status scores into status levels by splitting the scores into $K$ equal size bins; and then we assume nodes in the same status level share the same status ranking. Instead of rankings $1$ to $N$, now we have $K$ ranking levels. Then the global ranking constraint is relaxed to preserve status levels of nodes. 

To maintain the order between the nodes from different levels, we randomly generate a set of lists $\mathcal{L}={\{l_i\}}_{i=1}^{L}$. Each list $l=\{v_{[1]},v_{[2]}\dots,v_{[K]}\}$ consists of $K$ nodes, one from each level and the subscript of $v_{[i]}$ represents the level and thus the ranking information of the node. If we preserve the ranking levels of all individual lists, we can preserve the status level globally.  Therefore, we assume that all the lists are independent, and the probability the ranking of all the lists hold can be modeled as:
{\small
\begin{align}
p_{global} =\prod\limits_{l\in \mathcal{L}} p(l);
\label{eq:p_global_new}
\end{align}
}
where $p(l)$ is the probability that the order in list $l$ is maintained, which is defined as
{\small
\begin{align}
p(l) = \prod\limits_{1\leq i <j\leq K} p(v_{[i]},v_{[j]})
\end{align}
}
With this relaxation, the objective function that we want to minimize can be formalized as follows
{
\small
\begin{align}
\label{eq:optimization_global2}
O_1(\mathbf{U},{\bf w}) &= -\sum\limits_{l\in\mathcal{ L}}\sum\limits_{1\leq i <j\leq K}\log(p(v_{[i]},v_{[j]})) \\
% &=  -\sum\limits_{l\in L} \sum\limits_{1\leq i <j\leq K}\log (\sigma(f({\bf u}_{[i]}) - f({\bf u}_{[j]})))\nonumber\\
&= -\sum\limits_{l\in \mathcal{L}} \sum\limits_{1\leq i <j\leq K}\log (\sigma({\bf w}^T{\bf u}_{[i]} - {\bf w}^T{\bf u}_{[j]})) \nonumber \\
& := -\sum\limits_{l\in \mathcal{L}}O_l(\mathbf{U}_l,{\bf w});\nonumber
\end{align}
}
where $\mathbf{U}_l$ is representations for the nodes in list $l$.

%The optimization problem can be formalized as follows
%\begin{align}
%\mathcal{U}, {\bf w} &=\arg\min\limits_{\mathcal{U},{\bf w}} O(\mathcal{U},{\bf w}) 
%\label{eq:optimization_global2}
%\end{align}
To optimize \eqref{eq:optimization_global2}, we adopt the Stochastic Gradient Decent method. The procedure of the optimization is described in Algorithm \ref{alg:GSPE}. The inputs of this algorithm are the size of the set of random list $L$, the dimension of the representation $d$ and the learning rate $\eta$. Instead of generating the whole random list set beforehand, we generate a list $l$ in each step as the procedure proceeds in Line 5. We then calculate the derivatives for ${\bf w}$ and $\mathbf{U}_l$ and update them by Gradient Descent in Line 6. The generation of the list costs $O(K)$ time. For each list $l$, there are $K\cdot(K-1)/2$ pairs, which takes $O(d\cdot K^2)$ time to perform the update, thus, the whole procedure costs $O(L\cdot d\cdot K^2)$. 
\begin{algorithm}[!ht]
\caption{Optimization Procedure For GINE}
\label{alg:GSPE}
\begin{algorithmic}[1]
\STATE{\textbf{Input:} $L,d, \eta$}
\STATE{Initialize $d$-dimension representations $\mathbf{U}$ and $d$-dimension parameter ${\bf w}$.}
\STATE{$i\leftarrow 0$}

\WHILE{$i\leq L$}
\STATE{Generate a list $l=\{v_{[1]},v_{[2]}\dots,v_{[K]}\}$, which contains one node from each level.}
%\STATE Calculate the derivatives for ${\bf u}_{[j]}, j=1\dots K$ and ${\bf w}$ as follows
%\begin{align*}
%&\frac{\partial O_l(\mathcal{U}_l,{\bf w})}{\partial {\bf u}_{[j]}},\quad j=1,\dots K\\
%&\frac{\partial O_l(\mathcal{U}_l,{\bf w})}{\partial {\bf w}}
%\end{align*}
\STATE Update the representations and parameters by Gradient Descent as follows
{\small
\begin{align*}
& \mathbf{U}_l \leftarrow \mathbf{U}_l - \eta\cdot \frac{\partial O_l(\mathbf{U}_l,{\bf w})}{\partial \mathbf{U}_l}\\
&{\bf w}\leftarrow {\bf w} - \eta \cdot \frac{\partial O_l(\mathbf{U}_l,{\bf w})}{\partial {\bf w}}
\end{align*}
}
\STATE $i\leftarrow i+1$
\ENDWHILE
\RETURN $\mathbf{U},{\bf w}$
\end{algorithmic}
\end{algorithm}

\subsection{Local and Global Information Preserved Embedding.} In this subsection, we first briefly describe the local information preserved embedding model and then introduce the proposed framework LOG, which can learn node representation preserving both local and global information. 

\subsubsection{Preserving Local Information.}

To learn node representation that can preserve local information, we follow word2vec \cite{mikolov2013distributed}\cite{mikolov2013efficient}, which is an effective and efficient way to learn distributed representations for words. Many network embedding methods including LINE and DeepWalk are inspired by this model. The skip-gram model \cite{mikolov2013distributed}\cite{mikolov2013efficient} proposed in word2vec predicts surrounding context words given a center word, which can be formulated as follows:
\begin{align}
p(N(w_c)|w_c),
\label{eq:skipgram}
\end{align}
where $N(w_c)$ is the set of words that surround word $w_c$. The word representations can be learned by modeling \eqref{eq:skipgram} using the representations and maximizing it with respect to the representations.

In a similar way, we can view the nodes in a network as the ``words''. Then, for a node $v$, we can regard all the nodes connected to $v$ as the surrounding context ``words'', which can be denoted as $N(v)$.

Assuming the independence between observing different nodes given a center node $v$, the probability that $N(v)$ is the surrounding ``context'' of node $v$ can be modeled as:
{\small
\begin{align}
p(N(v)|v) = \prod \limits_{v_j\in N(v)} p(v_j|v);
\label{eq:1}
\end{align}
}
where $p(v_j|v)$ can be modeled using a softmax function as suggested in \cite{mikolov2013distributed}.
{\small
\begin{align}\label{eq:softmax}
p(v_j|v) = \frac{\exp({\bf u}^T {{\bf u}'}_j)}{\sum\limits_{v_i \in \mathcal{V}}\exp({\bf u}^T {{\bf u_i}'})};
\end{align}
}
where $\mathcal{V}$ is the entire set of nodes in the network and ${\bf u}$ and ${\bf u}'$ are the source and target representations for node $v$ as suggested in \cite{mikolov2013distributed}.
 
To learn the representations for all the nodes in the network, we model this problem as a maximum likelihood problem. In other words, we need to maximize the probability that $N(v)$ is the ``context'' of node $v$ for all the nodes $v\in \mathcal{V}$ with respect to the node representations:
{\small
\begin{align}
P_{local} = \prod \limits_{v\in \mathcal{V}} p(N(v)|v).
\label{eq:p_local}
\end{align}
}
The objective function to be minimized is the negative logarithm of \eqref{eq:p_local}
{\footnotesize
 \begin{align}
 \label{eq:optimization_local}
 O_2(\mathbf{U}) &=  -\sum \limits_{v\in \mathcal{V}} \sum \limits_{v_j\in N(v)} \log p(v_j|v) \\
 &=-\sum \limits_{v\in \mathcal{V}} \sum \limits_{v_j\in N(v)} \log \frac{\exp({\bf u}^T {{\bf u}'}_j)}{\sum\limits_{v_i \in \mathcal{V}}\exp({\bf u}^T {{\bf u_i}'})} \nonumber
 \end{align}
}
%Instead of directly maximizing \eqref{eq:p_local}, we minimize the negative logarithm of it. The representations can be obtained by minimizing the following objective function
%\begin{align}
%O_1& =-\log P_{local} \nonumber\\
%&=-\sum\limits_{v \in \mathcal{V}} \log p(N(v)|v)\label{eq:objective1} \\
%&= -\sum\limits_{v \in \mathcal{V}} \sum\limits_{v_j \in N(v)} \log p(v_j|v) \nonumber 
%\end{align}
\subsubsection{The LOG Framework.}
To learn the representations preserving both the local and global information, the global information needs to be incorporated into the local information preserved embedding model. We combine the two objective \eqref{eq:optimization_local} and \eqref{eq:optimization_global2} as:
%\begin{align}
%P = p_{local}\cdot p_{global}^{\lambda}
%\label{eq:p_combine} 
%\end{align}
{\small
\begin{align}
&O(\mathbf{U},\mathbf{U}',{\bf w},{\bf w}')= \label{eq:objective_combine1}\\
& -\sum \limits_{v\in \mathcal{V}} \sum \limits_{v_j\in N(v)} \log \frac{\exp({\bf u}^T {{\bf u}'}_j)}{\sum\limits_{v_i \in \mathcal{V}}\exp({\bf u}^T {{\bf u_i}'})};\nonumber\\
& -\lambda \sum\limits_{l\in L} \sum\limits_{1\leq i <j\leq K}\log (\sigma(f({\bf u}_{[i]},{\bf u}'_{[i]})-f({\bf u}_{[j]},{\bf u}'_{[j]}) )\nonumber
\end{align}
}
where $\lambda\in [0,1]$ is a hyperparameter which indicates the importance of the global information.

The node representations $\mathbf{U},\mathbf{U}'$ and the parameters $\mathbf{w},\mathbf{w}'$ can be obtained by minimizing the objective function \eqref{eq:objective_combine1}. Note that, the global part of this objective function can be used to extend some existing network embedding methods such as LINE to incorporate the global information.

Since for each node, we have two representations, hence, we need to slightly modify the mapping function $f$ as 
{\small
\begin{align}
f({\bf u},{\bf u}') = {\bf w}^T{\bf u} + {\bf w}'^T{\bf u}'
\end{align}
}
After we obtain the source and target representations $\mathbf{U}$, $\mathbf{U}'$, we concatenate them to generate the final node representations. 

\subsubsection{An Optimization Method for LOG.}
\vspace{-2mm}
To optimize \eqref{eq:objective_combine1}, we adopt Stochastic Gradient Descent method. We view the two terms as two optimization tasks. We regard the local information task as the main task which will always be performed and the global status task will be performed with probability $\lambda$. The optimization of the global status task has been discussed in subsection \ref{sec:optimization_global}, so, we only discuss the optimization of the local term and the joint optimization of the two tasks.

We observe that the minimization of the first term of \eqref{eq:objective_combine1} is computational expensive due to the summation over the whole set of nodes $\mathcal{V}$ in \eqref{eq:softmax}. Thus, we adopt the negative sampling approach proposed in \cite{mikolov2013distributed} to solve the computational issue in \eqref{eq:softmax}. By using the negative sampling method, we replace each $\log p(v_i|v)$ with
{\small
\begin{align}
g(v,v_i) = \log\sigma({\bf u}^T {{\bf u}'}_i) + \sum\limits_{n=1}^{N_e} \log \sigma(-{\bf u}^T {{\bf u}'}_{\{n\}});
\end{align}
}
where $\sigma(x)$ is the sigmoid function and $Ne$ is the number of negative samples. The negative samples are sampled from the node set $\mathcal{V}$ according to the noise distribution $P(v)\sim d_v^{3/4}$ as proposed in \cite{mikolov2013distributed}, where, $d_v$ is the in-degree of the node $v$. The objective function becomes
{\small
\begin{align}
\label{eq:objective_combine2}
&O(\mathbf{U},\mathbf{U}',{\bf w},{\bf w}')= \\
& -\sum \limits_{v\in \mathcal{V}} \sum \limits_{v_i\in N(v)} \log\sigma({\bf u}^T {{\bf u}'}_i) + \sum\limits_{n=1}^{N_e} \log \sigma(-{\bf u}^T {{\bf u}'}_{\{n\}}) \nonumber\\
& -\lambda \sum\limits_{l\in \mathcal{L}} \sum\limits_{1\leq i <j\leq K}\log (\sigma(f({\bf u}_{[i]},{\bf u}'_{[i]})-f({\bf u}_{[j]},{\bf u}'_{[j]}) )\nonumber\\
&:= -\sum \limits_{v\in \mathcal{V}} \sum \limits_{v_i\in N(v)} g(v,v_i)\nonumber - \lambda \sum\limits_{l\in \mathcal{L}}O_l(\mathbf{U}_l, \mathbf{U}'_l,{\bf w},{\bf w}')\nonumber
\end{align}
}
The joint optimization procedure for LOG is described in Algorithm \ref{alg:LOG}. The inputs of this algorithm are the representation dimension $d$, the learning rates for the two tasks $\eta_1$ and $\eta_2$, the hyperparameter $\lambda$ controlling the importance of the global term, the number of negative samples $N_e$ and the number of running epochs $E$. The representations $\mathbf{U,U}'$ and the parameters $\mathbf{w,w}'$ are initialized in line 2. The algorithm goes through the entire node set $\mathcal{V}$ $E$ times. In the start of each epoch, the node set is shuffled to achieve better performance of SGD. Lines $7$-$10$ perform the update of the representations corresponding to the local term of \eqref{eq:objective_combine2}. Lines $11$-$15$ perform the update of the representations and parameters corresponding to the global term of \eqref{eq:objective_combine2} with probability $\lambda$. The update procedure of the local term in lines $8$-$9$ takes $O(d\cdot N_e)$ time, while the update procedure for the global term in lines $11$-$15$ takes $O(d\cdot K^2)$ time. As the for loop in line $6$ goes through all the nodes and the for loop in line $7$ goes through all the edges connected to the node, each epoch of procedure take $O(M\cdot d \cdot N_e + N \cdot d \cdot K^2 )$ time, where $M$ is the number of edges and $N$ is the number of nodes in the network. Therefore, the whole procedure costs $O(E\cdot( M\cdot d \cdot N_e + N \cdot d \cdot K^2) )$.

\begin{algorithm}[!h]
\caption{The Optimization Procedure for LOG}
\label{alg:LOG}
\begin{algorithmic}[1]
\STATE{\textbf{Input:} $d, \eta_1, \eta_2,\lambda, N_e,E$}
\STATE{Initialize $d$-dimension representations $\mathbf{U}$,$\mathbf{U}'$ and $d$-dimension parameter ${\bf w},$ ${\bf w}'$.}
\STATE{$i \leftarrow 0$}
\WHILE{$i\leq E$}
\STATE{Shuffle($\mathcal{V}$)}
\FOR{$v\in \mathcal{V}$}
\FOR{$v_j \in N(v)$}
\STATE{Generate $N_e$ negative samples}.
\STATE{Update the representations by Gradient Decent as follows}
%\begin{align*}
%&\mathbf{u} \leftarrow \mathbf{u} -\eta_1 \cdot \frac{\partial g(v,v_j)}{\partial \mathbf{u}};\\
%&\mathbf{u}_j \leftarrow \mathbf{u}_j -\eta_1 \cdot \frac{\partial g(v,v_j)}{\partial \mathbf{u}_j};\\
%&\mathbf{u}_{\{n\}} \leftarrow \mathbf{u}_{\{n\}} -\eta_1 \cdot \frac{\partial g(v,v_j)}{\partial \mathbf{u}_{\{n\}}}\quad n=1,\dots N_e.
%\end{align*}
{\small
\begin{align*}
&\mathbf{\Psi} \leftarrow \mathbf{\Psi} -\eta_1 \cdot \frac{\partial g(v,v_j)}{\partial \mathbf{\Psi}};
\end{align*}
}
where $\mathbf{\Psi}= \{\mathbf{u},\mathbf{u}'_j,\mathbf{u}'_{\{n\}},  n=1,\dots,N_e\}$ denotes all the involved representations in $g(v,v_j)$.
\ENDFOR
\STATE Generate a random number $p\in[0,1]$.
\IF{$p\leq \lambda$}
\STATE{Generate a list $l=\{v_{[1]},v_{[2]}\dots,v_{[K]}\}$, which contains one node from each status level.}

\STATE Update the representations and parameters by Gradient Decent as follows
%\begin{align*}
%& \mathbf{U}_l \leftarrow \mathbf{U}_l - \eta_2\cdot \frac{\partial O_l(\mathbf{U}_l,\mathbf{U}'_l,{\bf w},{\bf w}')}{\partial \mathbf{U}_l}; \\
%&\mathbf{U}'_l \leftarrow \mathbf{U}'_l - \eta_2\cdot \frac{\partial O_l(\mathbf{U}_l,\mathbf{U}'_l,{\bf w},{\bf w}')}{\partial \mathbf{U}'_l};\\ 
%&{\bf w}\leftarrow {\bf w} - \eta_2 \cdot \frac{\partial O_l(\mathbf{U}_l,\mathbf{U}'_l,{\bf w},{\bf w}')}{\partial {\bf w}};\\
%&{\bf w}'\leftarrow {\bf w}' - \eta_2 \cdot \frac{\partial O_l(\mathbf{U}_l,\mathbf{U}'_l,{\bf w},{\bf w}')}{\partial {\bf w}'}.
%\end{align*}
{\small
\begin{align*}
&\mathbf{\Phi} \leftarrow \mathbf{\Phi} -\eta_2 \cdot \frac{\partial O_l(\mathbf{U}_l,\mathbf{U}'_l,{\bf w},{\bf w}')}{\partial \mathbf{\Phi}};
\end{align*}
}
where $\mathbf{\Phi}= \{\mathbf{U}_l,\mathbf{U}'_l,{\bf w},{\bf w}'\}$ denotes all the involved representations and parameters in $O_l(\mathbf{U}_l,\mathbf{U}'_l,{\bf w},{\bf w}')$.
\ENDIF
\ENDFOR
\STATE{$i\leftarrow i + 1$}
\ENDWHILE
\RETURN $\mathbf{U},\mathbf{U}',{\bf w},{\bf w}'$
\end{algorithmic}
\end{algorithm}

\section{Experiments}
\label{sec:experiments}

In this section, we conduct experiments to verify the effectiveness of the proposed algorithms GINE and LOG. We first show that the embeddings learned by GINE and LOG can preserve the global information. Then, we perform link prediction to verify that including the global information can help learn better representations.  To evaluate the effectiveness of the proposed model, we conduct experiments on two network datasets, which have been widely used to evaluate network embedding algorithms~\cite{perozzi2014deepwalk,wang2016structural}.
\begin{itemize}
\item {\bf BlogCatalog}~\cite{Zafarani+Liu:2009} is a network of social relationships provided by blogger authors. This network consists of $10,312$ nodes and $333,983$ edges. 
\item {\bf Flickr}~\cite{Zafarani+Liu:2009} is a network of contacts between users from the photo sharing website Flickr. This network consists of $80,513$ nodes and $5,899,882$ edges.
\end{itemize}
In our experiments, we empirically set the number of levels $K$ to $60$. 
\subsection{Validating Global Information Preserving.}
\begin{figure*}[!ht]
\centering
  \begin{subfigure}[b]{0.47\linewidth}
    \includegraphics[width=0.9\linewidth]{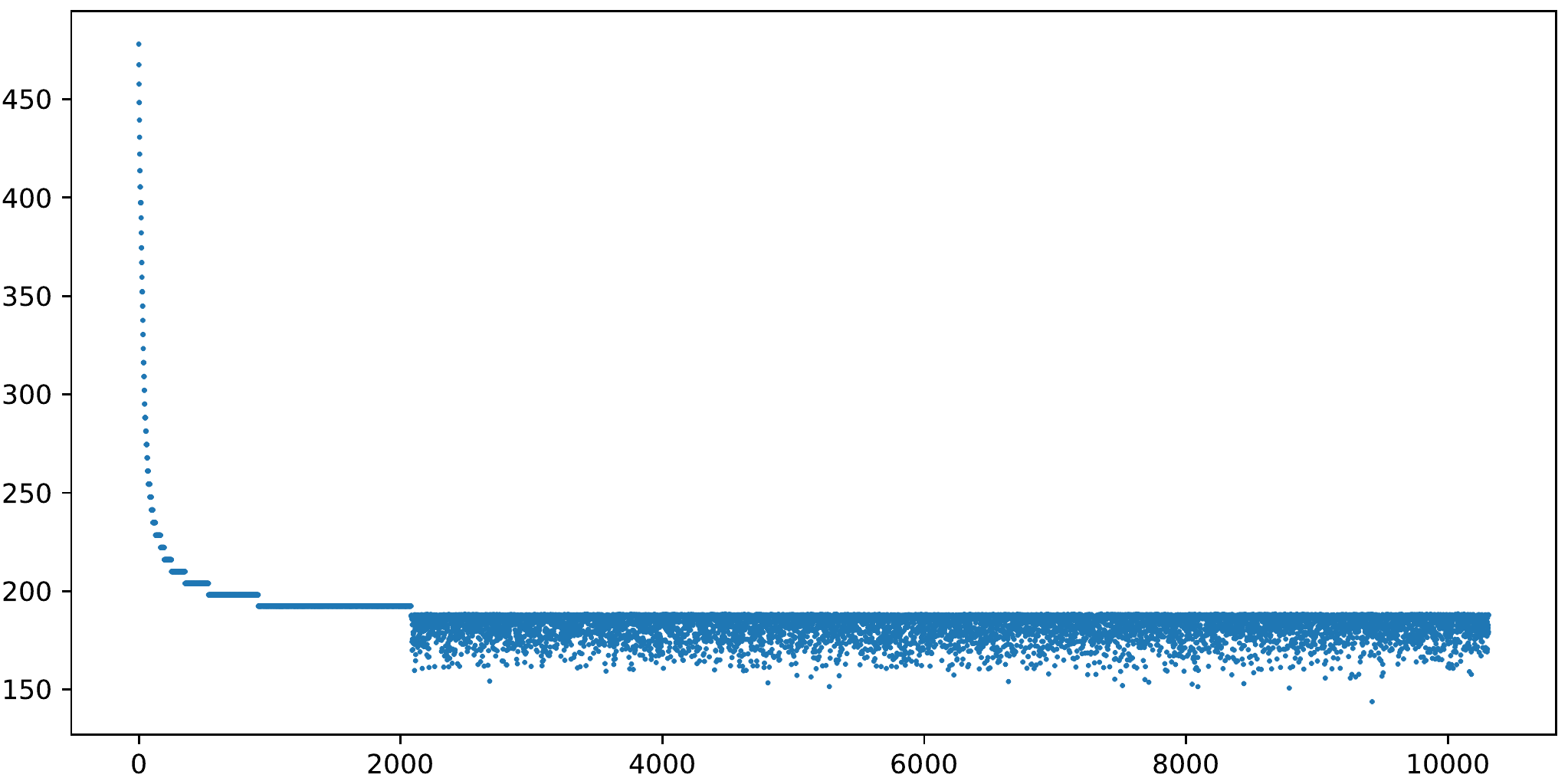}
    \caption{GINE}
    \label{fig:global}
  \end{subfigure}
   \begin{subfigure}[b]{0.47\linewidth}
    \includegraphics[width=0.9\linewidth]{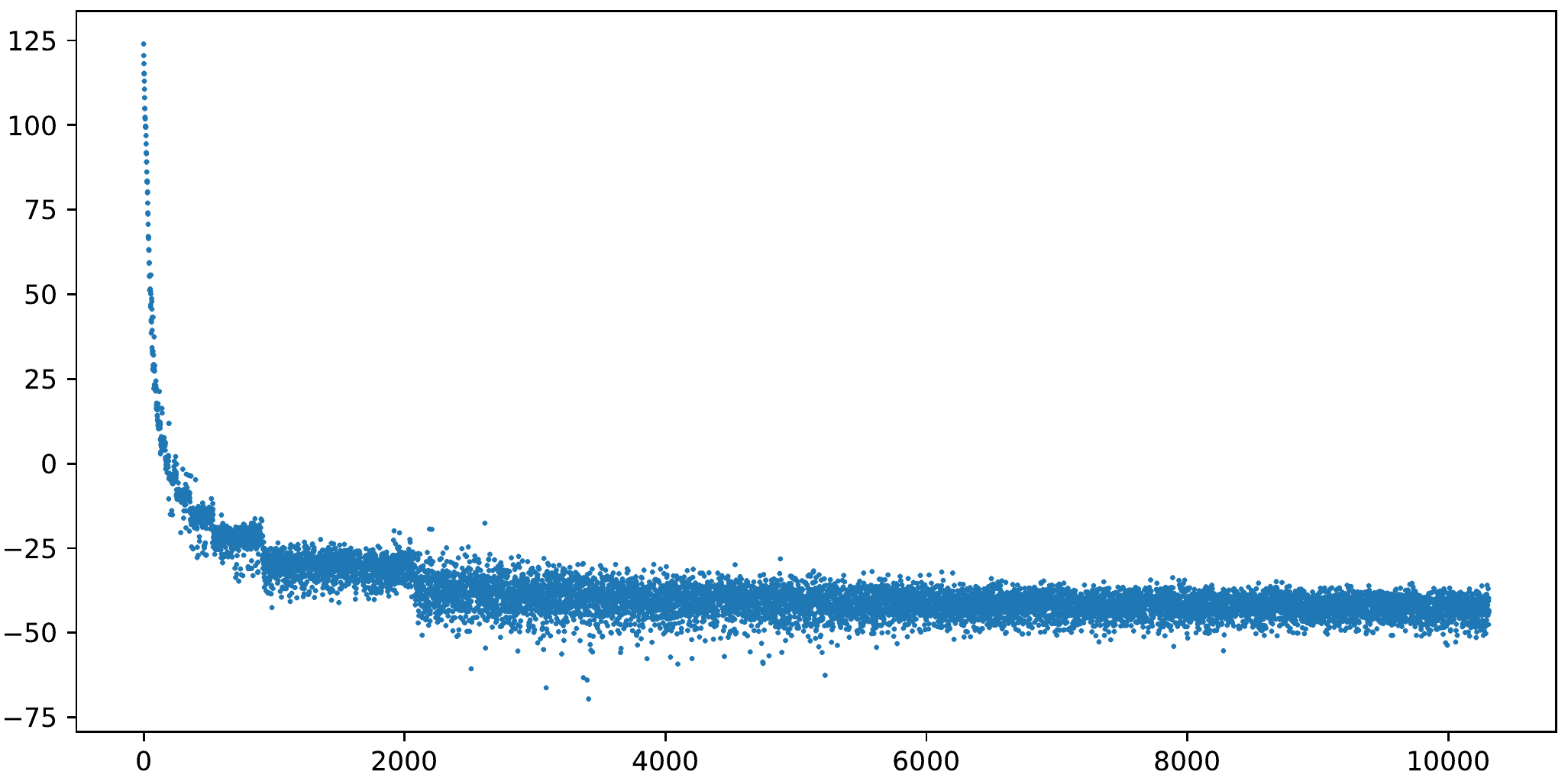}
    \caption{LOG with $\lambda=0.5$}
    \label{fig:global_local}
  \end{subfigure}
      \vskip\baselineskip
    \begin{subfigure}[b]{0.47\linewidth}
    \includegraphics[width=0.9\linewidth]{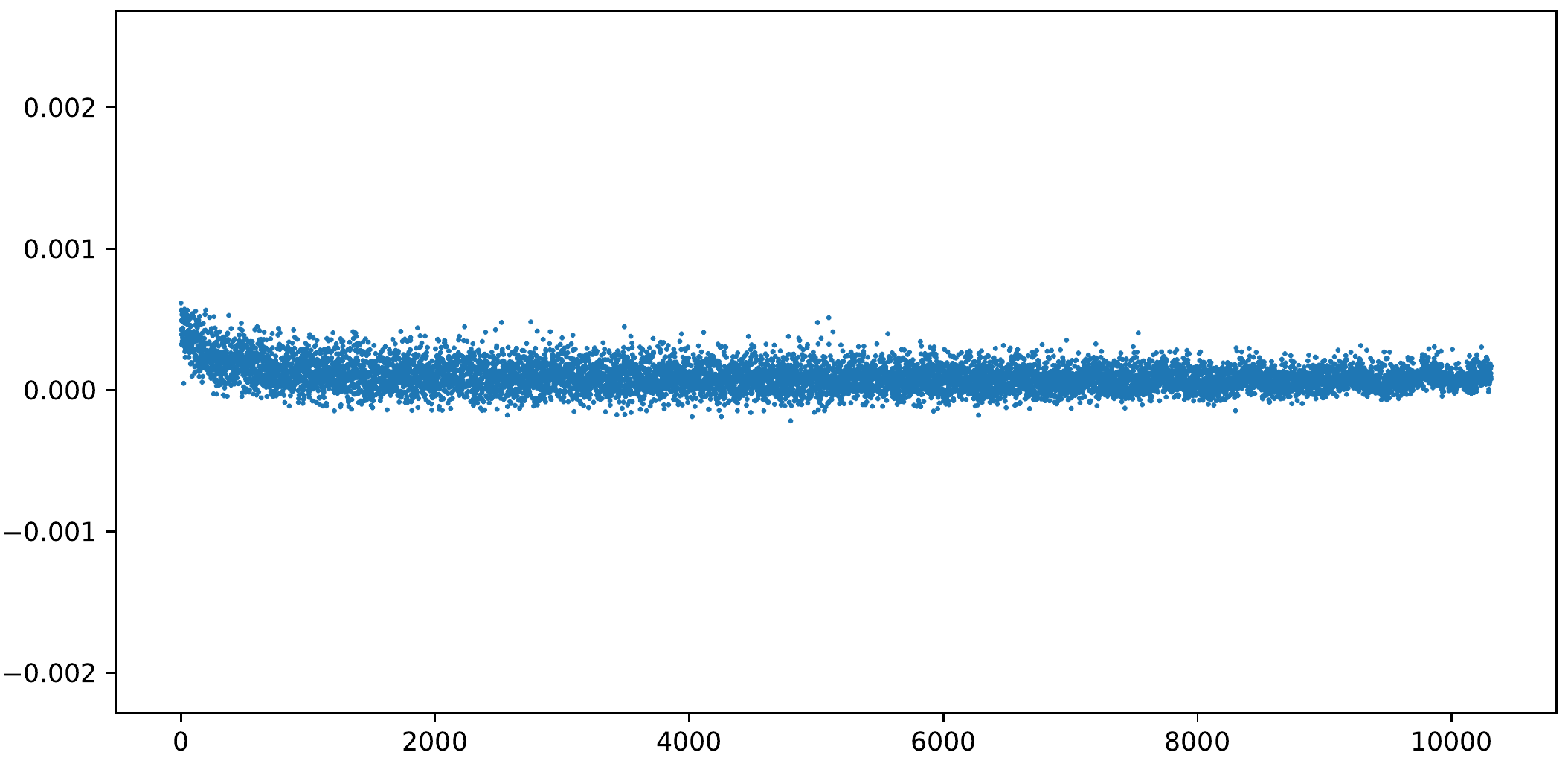}
    \caption{ DeepWalk}
    \label{fig:deepwalk}
  \end{subfigure}
      \begin{subfigure}[b]{0.47\linewidth}
    \includegraphics[width=0.9\linewidth]{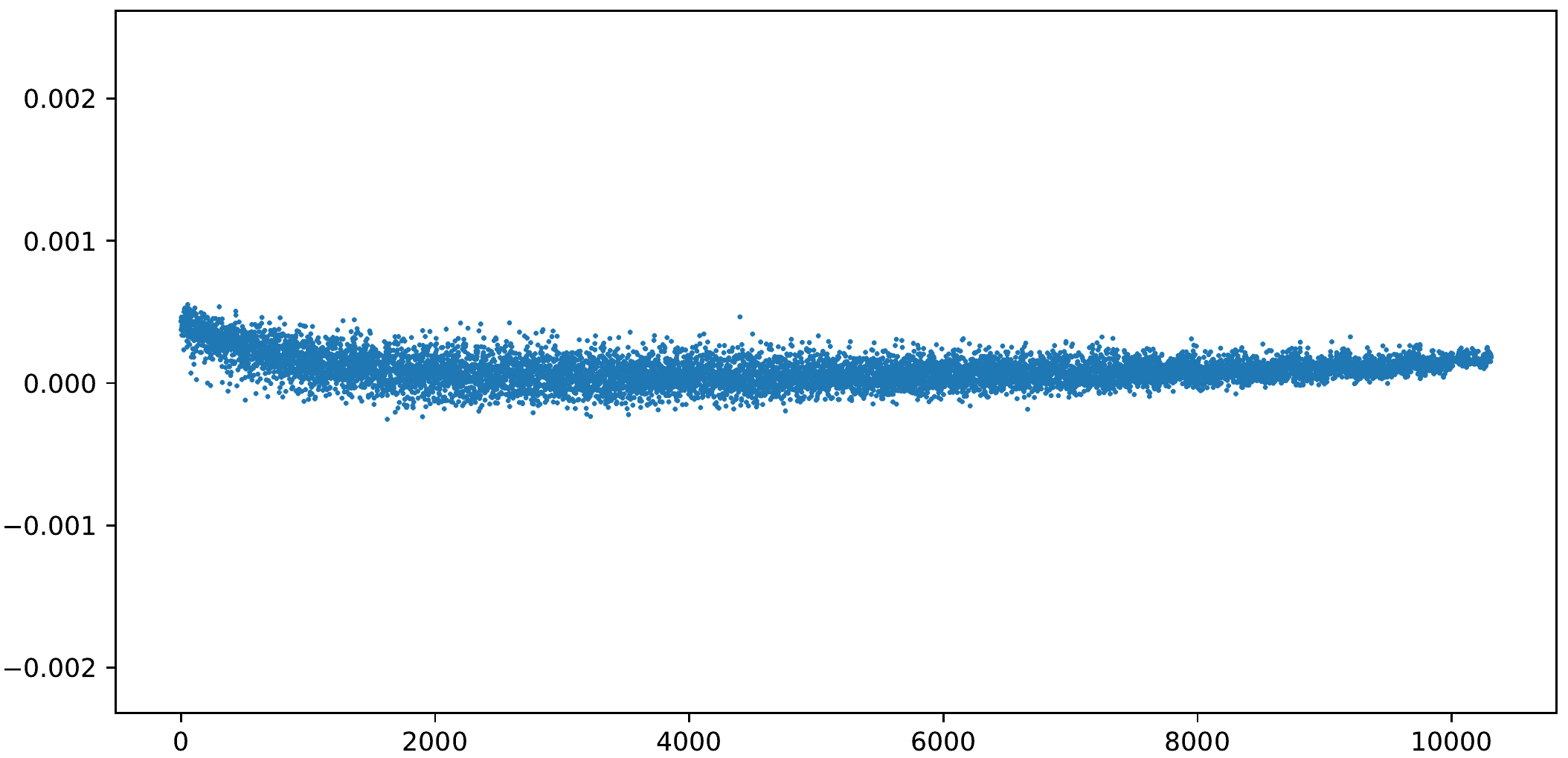}
    \caption{LINE}
    \label{fig:LINE}
  \end{subfigure}
  \vspace{-2mm}
  \caption{Global Status Preserved by Different Methods}
  \label{fig:global_status}
  \vspace{-4mm}
\end{figure*}

In this subsection, we show that the representations learned by our algorithms can preserve the global information. To visualize the global status of all the nodes in the network, we plot scatter plots for each method. For the BlogCatalog network, the scatter plot of the PageRank score has been shown in Figure~\ref{fig:blog_pagerank}. In this figure, the nodes are arranged in a descending order based on the PageRank score. For the plots of the other algorithms, we also arrange the nodes in this order. Figures~\ref{fig:global} and~\ref{fig:global_local} show the global status preserved by the proposed methods GINE and LOG respectively. Figure~\ref{fig:deepwalk} and~\ref{fig:LINE} show the global status preserved by DeepWalk and LINE respectively. DeepWalk and LINE do not learn a mapping function during the representation learning, therefore, we perform a linear regression using the learned representations and the original PageRank scores and then use this linear model to calculate the global status for each node. Note that since our models GINE and LOG use a linear mapping function from representations to the status scores, we also use a linear mapping function for DeepWalk and LINE for a fair comparison. 

We only show the results in the BlogCatalog network in Figure~\ref{fig:global_status} as the performance in the Flickr network is similar. As shown in Figure~\ref{fig:global}, GINE can preserve the global status quite well, and the difference between nodes in different levels are clear. LOG can also roughly preserve the ranking information. Although it is noisier than GINE, the trend has been preserved, and the difference between levels can also be observed clearly, which achieves our goal of loosely preserving the global ranking. Figure~\ref{fig:deepwalk} and Figure~\ref{fig:LINE} do not show a specific trend, which means the representations learned by DeepWalk and LINE cannot preserve the global information. 

\subsection{Link Prediction.}
In this subsection, we perform link prediction task to further evaluate the effectiveness of the learned representations. The intuition is that a better embedding algorithm should learn better node representations, which will lead to better link prediction performance. 

\subsubsection{Experiments Setting.}
In the link prediction task, a certain fraction of edges are removed from the network, we are supposed to use the remained network to predict whether the ``missing'' edges exist. 

We perform the link prediction on the aforementioned two datasets. For each dataset, we set up three groups of experiments where $20\%,50\%$ and $80\%$ of edges are removed, while at least one edge for each node will remain in the network. After removing the edges, we use the remained network to learn the node representations. Then, to perform link prediction, we need to form representations for an edge or a pair of nodes. We use the element-wise addition of the two node representations as the representation of the edge. 
 
To form the training set, we put all the edges remained in the network as positive samples and then sample an equal number of non-connected node pairs as negative samples. The testing set is formed in the same way.  After forming the training set and testing set, we train a binary classifier using logistic regression on the training set and perform link prediction on the testing set. In this work, we use accuracy and AUC as the metrics to evaluate the link prediction performance. 

\begin{table*}[h!]
 \centering
  \begin{tabular}{c|c|ccc|ccc}
    \toprule
    \multicolumn{2}{c}{DataSet}&\multicolumn{3}{|c}{BlogCatalog}&\multicolumn{3}{|c}{Flickr}\\
    \midrule
    &$\%$ removed edges &$20\%$ & $50\%$& $80\%$  &$20\%$ & $50\%$ & $80\%$ \\
    \toprule
   \multirow{6}{*}{Accuracy}& LOG(0.3) & {\bf 87.93}& {\bf 87.46}& {\bf 88.16} & {\bf 83.46}& {\bf 83.00}&{\bf 83.75}\\
    &LINE & 86.66 & 86.44&85.93 & 80.12& 80.45&82.31\\
  %  LINE(c)&  & &\\
    &DeepWalk& 86.40& 85.98& 83.56 &80.92& 80.56 &71.37\\
  %  Deepwalk(c)& & &\\
    &HFB & 76.96 &72.54 &61.27&75.73&70.69&62.73 \\
    &GINE& 84.84&83.48&83.39 &78.74&79.57&80.43\\
    &LOG(0)&86.89&85.70&85.80 &80.84 &80.38 & 81.74\\
    \bottomrule
    
   % \midrule
%   \multirow{6}{*}{Macro-$F_1(\%)$}& GlobalLocal & {\bf 87.84}& {\bf 87.09}&{\bf 88.16}  &82.63 & 82.99& 83.75 \\
%    &LINE & 86.66& 86.43&85.93& 79.64&80.35 & 82.31\\
%  %  LINE(c)&  & &\\
%    &DeepWalk& 86.39&  85.98&83.55 \\
%  %  Deepwalk(c)& & &\\
%    &FeatureBased  &  76.89 & 71.16 &54.52&74.61&68.19&56.84  \\
%    &Global&84.75 &83.39 &83.31&78.65&79.48&80.37 \\
%    &Local&86.89&85.70&85.80 & 80.84& 80.37& 81.73\\
%    \bottomrule
%    
       % \midrule
   \multirow{4}{*}{AUC}& LOG(0.3) &{\bf 0.9473} & {\bf 0.9445}& {\bf 0.9491} &{\bf 0.9113} &{\bf 0.9061} &{\bf 0.9115}\\
    &LINE &  0.9398&  0.9316&0.9284 & 0.8805& 0.8962&0.9064  \\
  %  LINE(c)&  & &\\
    &DeepWalk& 0.9382& 0.9341&0.9139 &0.8896 &0.8856 &0.7836\\
  %  Deepwalk(c)& & &\\
    &HFB & 0.8693&0.8712  &0.8249&0.8635&0.8339&0.7852 \\
    &GINE& 0.9193&0.9016 &0.9007&0.8591&0.8679&0.8746\\
    &LOG(0)&0.9387&0.9292&0.9275& 0.8881& 0.8831& 0.8883 \\
    \bottomrule   
  \end{tabular}
 \vspace{0mm}
  \caption{Link Prediction Performance} 
  \label{tab:linkprediction}
  \vspace{-4mm}
\end{table*}

\subsubsection{Performance Comparison.}

To evaluate the performance of our algorithms, we compare them with the following representative baselines:
\begin{itemize}
\item \textbf{LINE}~\cite{tang2015line}: LINE is a network embedding algorithm which can preserve the first-order and second order proximity. We apply LINE to the networks and obtain the node representations.  
\vspace{-2mm}
\item \textbf{DeepWalk}~\cite{perozzi2014deepwalk}: DeepWalk facilitates local information obtained from random walk to perform network embedding. We apply DeepWalk to the networks and obtain the node representations.
\vspace{-2mm} 
\item \textbf{Handcrafted Feature Based(HFB)}: Given a pair of nodes, we calculate their number of common neighbors, Jaccard coefficient and preferential attachment value~\cite{liben2007link} as the representations for this node pair.
\vspace{-2mm}
\item \textbf{GINE}: This algorithm only uses the global information to learn the node representations.
\vspace{-2mm}
\item \textbf{LOG(0)}: This is a variant of our method LOG. We set $\lambda=0$ so that only local information is used to learn the node representation. 
\end{itemize}

For all the methods except for HFB, we set the dimension to $128$. The experiment results are shown in Table \ref{tab:linkprediction}. For LOG, we set $\lambda=0.3$, so we denote it as LOG(0.3) in Table~\ref{tab:linkprediction}. 
%All these parameters are set via cross validation. 
The following observations can be made from the results
\begin{itemize}
\item All embedding algorithms obtain better performance than HFB (Handcrafted Feature Based), which supports the advantage of the automated representation learning algorithms for networks.
\vspace{-2mm} 
\item Although the performance of GINE is not as good as the other embedding based methods, it provides reasonable results. This shows the importance and effectiveness of the global information.
%This means that preserving the global information can obtain better presentations for link prediction.
\vspace{-2mm}
\item The performance of LOG(0) is close to the performance of LINE, which is reasonable as LOG(0) is similar to LINE.
\vspace{-2mm}
\item LOG(0.3) consistently outperforms LOG(0), which further proves the importance of the Global information.
\vspace{-2mm} 
\item Our method LOG(0.3) outperforms all the baselines in each setting of the two datasets, which shows that (1) local and global information are complementary; and (2) incorporating them can lead to better presentation learning. 
\end{itemize}

\subsubsection{Parameter Analysis.}

In this subsection, we analyze how the parameters in our model affect the performance of the link prediction task. There are mainly two parameters in our model, the dimension of the representation $d$ and the hyperparameter $\lambda$ controlling the importance of the global information. We first analyze the stability of the performance of the link prediction task as the dimension $d$ changes. Then, we analyze how the performance changes as $\lambda$ varies.

\subsubsection{Stability over dimensions.}
\begin{figure}[!ht]
 \centering
    \begin{subfigure}[b]{0.45\linewidth}
    \centering
    \includegraphics[width=\linewidth]{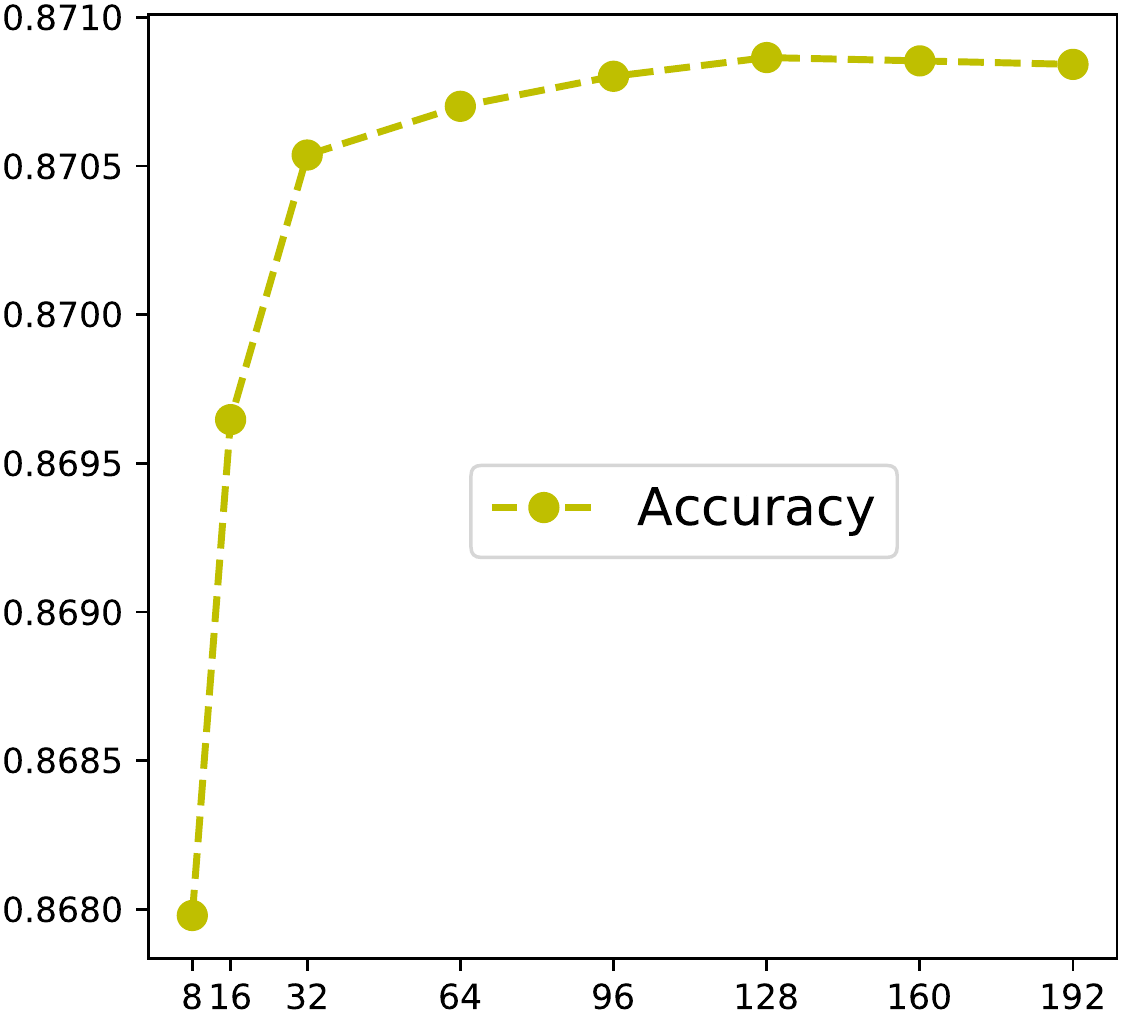}
    \caption{Accuracy}
    \label{fig:accuracy_dimensions}
  \end{subfigure}
      \begin{subfigure}[b]{0.45\linewidth}
      \centering
    \includegraphics[width=\linewidth]{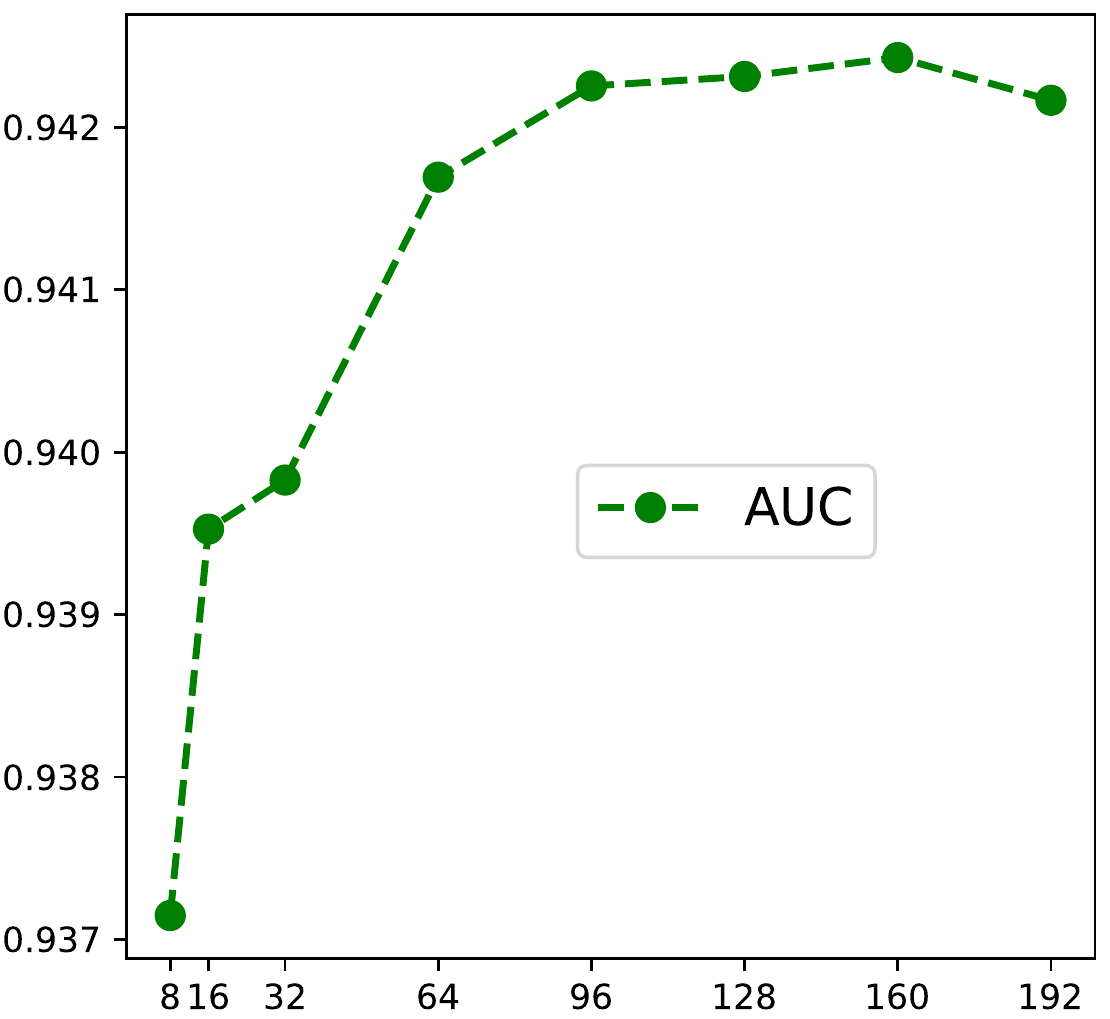}
    \caption{AUC}
    \label{fig:auc_diemsions}
  \end{subfigure}
  \caption{Different dimensions}
  \label{fig:parameter_analysis_dimensions}
\end{figure}

We analyze how the change of dimension affects the performance of the link prediction task. In particular, we set the dimension to $[8,16,32,96,128,160,192]$ with the setting of $50\%$ edges removed and $\lambda = 0.1$. The results are shown in Figure~\ref{fig:parameter_analysis_dimensions}, both the accuracy and AUC increase as the dimension gets larger at first and then when the dimension is larger than $96$, the performance starts to be stable. The performance is not very sensitive to the change of the dimension after it reaches $96$.

\subsubsection{Stability over $\lambda$ values.}
\begin{figure}[!ht]
\centering
  \begin{subfigure}[b]{0.45\linewidth}
  \centering
  \includegraphics[width=\linewidth]{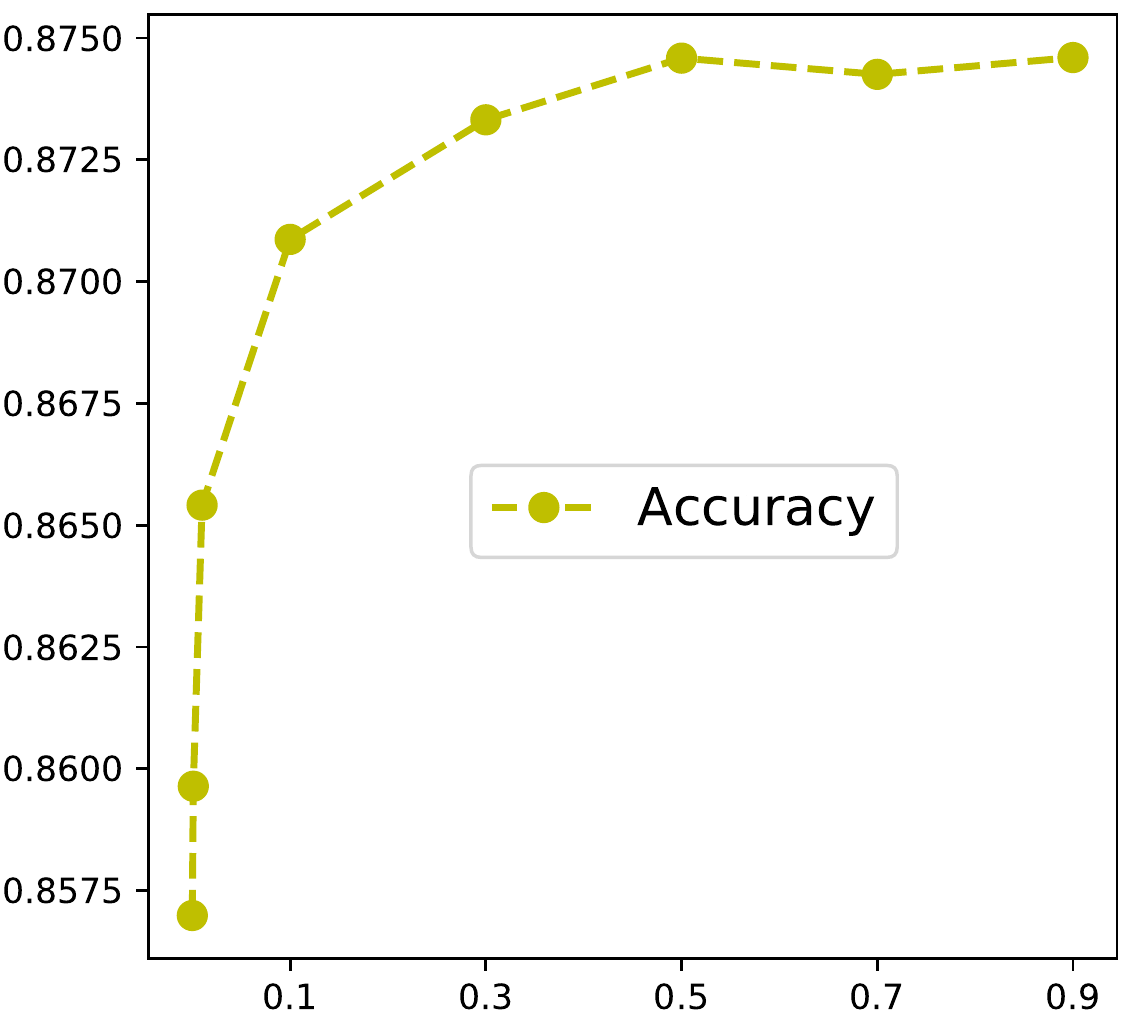}
    \caption{Accuracy}
    \label{fig:accuracy_lambda}
  \end{subfigure}
   \begin{subfigure}[b]{0.45\linewidth}
   \centering
    \includegraphics[width=\linewidth]{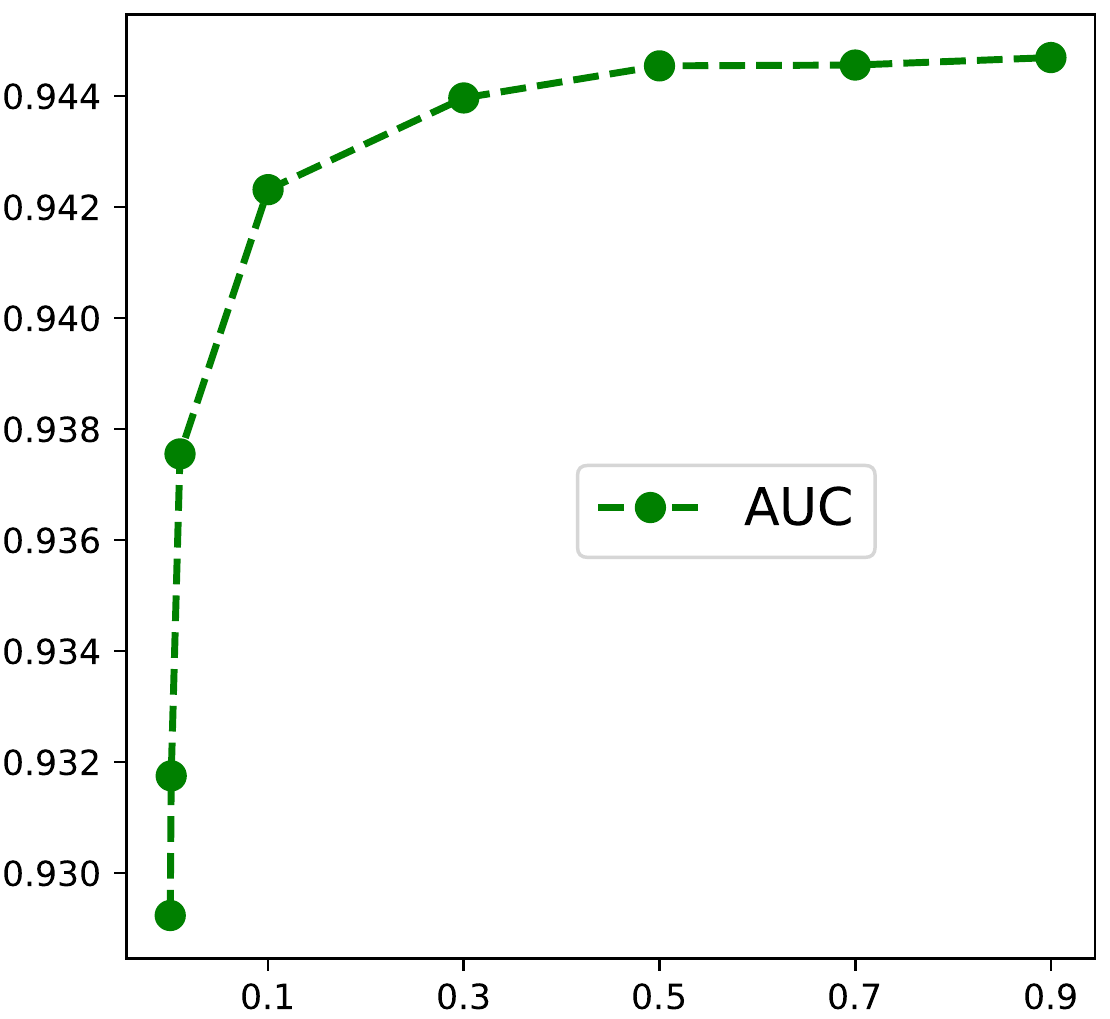}
    \caption{AUC}
    \label{fig:auc_lambda}
  \end{subfigure}
 %     \vskip\baselineskip
\vspace{-2mm}
 \caption{Stability over $\lambda$ values}
 \vspace{-4mm}
 \label{fig:parameter_analysis_lambda}
 \end{figure}

\vspace{-2mm}
We analyze how different values of $\lambda$ affect the link prediction performance. In particular, we set the $\lambda$ to $[0,0.001,0.01,0.1,0.3,0.5,0.7,0.9]$ with the setting of $50\%$ edges moved and dimension $128$. The results are shown in Figure~\ref{fig:parameter_analysis_lambda}. Both the accuracy and the AUC score increase as $\lambda$ gets large, which indicates the effectiveness of including global information. The performance starts to be stable when $\lambda$ is large or equal to $0.3$.

So the performance of link prediction is very sensitive to the change of $\lambda$ when it is small, which shows the effectiveness of the including global information. When the $\lambda$ is large, it is likely that enough global information has been absorbed in the representations and the performance cannot be improved by further increasing the value of $\lambda$.

\vspace{-2mm}
\section{Discussion and Future Work}
\label{sec:conclusions}
In this paper, we propose a principal way to model global information for network embedding. Based on this, we further propose a novel network embedding framework LOG, which can preserve both local and global information. The results of experiments show that LOG can well preserve the global information.The performance of link prediction further demonstrates that the inclusion of the global information can lead to better representation learning.  

In this paper, we use a linear function to model the global information. However, more complex functions can be adopted. The global status score we use in this paper is PageRank, while other global status scores could also be used. We will investigate these different possibilities in the future. 
%In this paper, we propose an approach to capture global information in network embedding. Our proposed approach preserves the global information by learning a mapping function an loosely maintain the global ranking of the status score while learning the node representations.  We also propose a framework to learn node representations which can preserve both global and local information. The results of experiments shows that our method can preserve the global information quite well in the sense of ranking. The results of link prediction further shows the effectiveness of the representations learned by our algorithm. 
\bibliographystyle{acm}
\bibliography{NEC} 

\begin{thebibliography}{10}

\bibitem{dong2017metapath2vec}
{\sc Dong, Y., Chawla, N.~V., and Swami, A.}
\newblock metapath2vec: Scalable representation learning for heterogeneous
  networks.
\newblock In {\em Proceedings of the 23rd ACM SIGKDD International Conference
  on Knowledge Discovery and Data Mining\/} (2017), ACM, pp.~135--144.

\bibitem{goyal2017graph}
{\sc Goyal, P., and Ferrara, E.}
\newblock Graph embedding techniques, applications, and performance: A survey.
\newblock {\em arXiv preprint arXiv:1705.02801\/} (2017).

\bibitem{grover2016node2vec}
{\sc Grover, A., and Leskovec, J.}
\newblock node2vec: Scalable feature learning for networks.
\newblock In {\em Proceedings of the 22nd ACM SIGKDD international conference
  on Knowledge discovery and data mining\/} (2016), ACM, pp.~855--864.

\bibitem{hamilton2017representation}
{\sc Hamilton, W.~L., Ying, R., and Leskovec, J.}
\newblock Representation learning on graphs: Methods and applications.
\newblock {\em arXiv preprint arXiv:1709.05584\/} (2017).

\bibitem{hu2013actnet}
{\sc Hu, X., Tang, J., Gao, H., and Liu, H.}
\newblock Actnet: Active learning for networked texts in microblogging.
\newblock In {\em Proceedings of the 2013 SIAM International Conference on Data
  Mining\/} (2013), SIAM, pp.~306--314.

\bibitem{huang2017label}
{\sc Huang, X., Li, J., and Hu, X.}
\newblock Label informed attributed network embedding.
\newblock In {\em Proceedings of the Tenth ACM International Conference on Web
  Search and Data Mining\/} (2017), ACM, pp.~731--739.

\bibitem{lee2011prioritizing}
{\sc Lee, I., Blom, U.~M., Wang, P.~I., Shim, J.~E., and Marcotte, E.~M.}
\newblock Prioritizing candidate disease genes by network-based boosting of
  genome-wide association data.
\newblock {\em Genome research 21}, 7 (2011), 1109--1121.

\bibitem{liben2007link}
{\sc Liben-Nowell, D., and Kleinberg, J.}
\newblock The link-prediction problem for social networks.
\newblock {\em journal of the Association for Information Science and
  Technology 58}, 7 (2007), 1019--1031.

\bibitem{massa2007survey}
{\sc Massa, P.}
\newblock A survey of trust use and modeling in real online systems.
\newblock In {\em Trust in E-services: Technologies, Practices and Challenges}.
  IGI Global, 2007, pp.~51--83.

\bibitem{mikolov2013efficient}
{\sc Mikolov, T., Chen, K., Corrado, G., and Dean, J.}
\newblock Efficient estimation of word representations in vector space.
\newblock {\em arXiv preprint arXiv:1301.3781\/} (2013).

\bibitem{mikolov2013distributed}
{\sc Mikolov, T., Sutskever, I., Chen, K., Corrado, G.~S., and Dean, J.}
\newblock Distributed representations of words and phrases and their
  compositionality.
\newblock In {\em Advances in neural information processing systems\/} (2013),
  pp.~3111--3119.

\bibitem{newman2005measure}
{\sc Newman, M.~E.}
\newblock A measure of betweenness centrality based on random walks.
\newblock {\em Social networks 27}, 1 (2005), 39--54.

\bibitem{okamoto2008ranking}
{\sc Okamoto, K., Chen, W., and Li, X.-Y.}
\newblock Ranking of closeness centrality for large-scale social networks.
\newblock {\em Lecture Notes in Computer Science 5059\/} (2008), 186--195.

\bibitem{page1999pagerank}
{\sc Page, L., Brin, S., Motwani, R., and Winograd, T.}
\newblock The pagerank citation ranking: Bringing order to the web.
\newblock Tech. rep., Stanford InfoLab, 1999.

\bibitem{perozzi2014deepwalk}
{\sc Perozzi, B., Al-Rfou, R., and Skiena, S.}
\newblock Deepwalk: Online learning of social representations.
\newblock In {\em Proceedings of the 20th ACM SIGKDD international conference
  on Knowledge discovery and data mining\/} (2014), ACM, pp.~701--710.

\bibitem{ramakrishnan2009mining}
{\sc Ramakrishnan, S.~R., Vogel, C., Kwon, T., Penalva, L.~O., Marcotte, E.~M.,
  and Miranker, D.~P.}
\newblock Mining gene functional networks to improve mass-spectrometry-based
  protein identification.
\newblock {\em Bioinformatics 25}, 22 (2009), 2955--2961.

\bibitem{ribeiro2017struc2vec}
{\sc Ribeiro, L.~F., Saverese, P.~H., and Figueiredo, D.~R.}
\newblock struc2vec: Learning node representations from structural identity.
\newblock In {\em Proceedings of the 23rd ACM SIGKDD International Conference
  on Knowledge Discovery and Data Mining\/} (2017), ACM, pp.~385--394.

\bibitem{tang2013exploiting}
{\sc Tang, J., Gao, H., Hu, X., and Liu, H.}
\newblock Exploiting homophily effect for trust prediction.
\newblock In {\em Proceedings of the sixth ACM international conference on Web
  search and data mining\/} (2013), ACM, pp.~53--62.

\bibitem{tang2016visualizing}
{\sc Tang, J., Liu, J., Zhang, M., and Mei, Q.}
\newblock Visualizing large-scale and high-dimensional data.
\newblock In {\em Proceedings of WWW\/} (2016), pp.~287--297.

\bibitem{tang2015pte}
{\sc Tang, J., Qu, M., and Mei, Q.}
\newblock {PTE:} predictive text embedding through large-scale heterogeneous
  text networks.
\newblock In {\em Proceedings of SIGKDD\/} (2015), pp.~1165--1174.

\bibitem{tang2015line}
{\sc Tang, J., Qu, M., Wang, M., Zhang, M., Yan, J., and Mei, Q.}
\newblock Line: Large-scale information network embedding.
\newblock In {\em Proceedings of the 24th International Conference on World
  Wide Web\/} (2015), International World Wide Web Conferences Steering
  Committee, pp.~1067--1077.

\bibitem{tang2010identifing}
{\sc Tang, X., and Yang, C.~C.}
\newblock Identifing influential users in an online healthcare social network.
\newblock In {\em Intelligence and Security Informatics (ISI), 2010 IEEE
  International Conference on\/} (2010), IEEE, pp.~43--48.

\bibitem{wang2016structural}
{\sc Wang, D., Cui, P., and Zhu, W.}
\newblock Structural deep network embedding.
\newblock In {\em Proceedings of the 22nd ACM SIGKDD international conference
  on Knowledge discovery and data mining\/} (2016), ACM, pp.~1225--1234.

\bibitem{wang2017signed}
{\sc Wang, S., Tang, J., Aggarwal, C.~C., Chang, Y., and Liu, H.}
\newblock Signed network embedding in social media.
\newblock In {\em Proceedings of SDM\/} (2017), pp.~327--335.

\bibitem{wang2016linked}
{\sc Wang, S., Tang, J., Aggarwal, C.~C., and Liu, H.}
\newblock Linked document embedding for classification.
\newblock In {\em Proceedings of CIKM\/} (2016), pp.~115--124.

\bibitem{Zafarani+Liu:2009}
{\sc Zafarani, R., and Liu, H.}
\newblock Social computing data repository at {ASU}, 2009.

\end{thebibliography}
\end{document}